\documentclass{rsproca}

\usepackage{bbm}
\usepackage{overpic} 
\usepackage{psfrag}
\newcommand\be{\begin{equation}}
\newcommand\ee{\end{equation}}
\newcommand\Rey{\mbox{\textit{Re}}}  
\providecommand\bnabla{\boldsymbol{\nabla}}
\providecommand\bcdot{\boldsymbol{\cdot}}
\def\00{\mathbf{0}}
\def\aa{\mathbf{a}}
\def\atan{\tan^{-1}}
\def\bb{\mathbf{b}}
\def\bdelta{\boldsymbol{\delta}}
\def\CC{\mathbf{C}}
\def\dd{\mathrm{d}}
\def\dumb{\gamma}
\def\ex{\mathbf{e}_x}
\def\ey{\mathbf{e}_y}
\def\nn{\mathbf{n}}
\def\nnu{\Rey^{-1}}
\def\Pa{P^\dag} 
\def\tt{\mathbf{t}}
\def\UU{\mathbf{U}}
\def\UUa{\mathbf{U}^\dag}
\def\xx{\mathbf{x}}

\begin{document}

\title{
Manipulating flow separation: sensitivity of 
stagnation points, 
separatrix angles and 
recirculation area to steady actuation
}

\author{
E. Boujo and F. Gallaire}

\address{Laboratory of Fluid Mechanics and Instabilities\\
\'Ecole Polytechnique F\'ed\'erale de Lausanne\\
CH-1015 Lausanne, Switzerland}

\subject{fluid mechanics}


\corres{E. Boujo\\
\email{edouard.boujo@epfl.ch}}

\begin{abstract}
A variational technique is used to derive analytical expressions for the sensitivity of several geometric indicators of flow separation to steady actuation. Considering the boundary layer flow above a wall-mounted bump, the six following  representative quantities are considered: the locations of the  separation point and reattachment point connected by the separatrix, the separation angles  at these stagnation points, the backflow area and the recircula-tion area. For each geometric quantity, linear sensitivity analysis allows us to identify regions which are the most sensitive to volume forcing and wall blowing/suction. Validations against full non-linear Navier$-$Stokes calculations show excellent agreement for small-amplitude control for all considered indica-tors. With very resemblant sensitivity maps, the reattachment point, the backflow and recirculation areas are seen to be easily manipulated. In contrast, the upstream separation point and the separatrix angles are seen to remain extremely robust with respect to external steady actuation. 
\end{abstract}


\maketitle

\section{Introduction}

Flow separation leads in many aerodynamic situations to performance loss, such as reduced lift, increased drag, enhanced fluctuations or noise production. 
In contrast, separation yields a recirculation region that is often desirable in combustion devices. 
It is thus not a surprise that there is extensive research on the control of flow separation \cite{Seifert06}. 
Attempts include, in decreasing order of complexity, closed-loop separation control, harmonic or steady active open-loop control and passive control devices.

Studies on closed loop control strategies remain few: while Alam, Liu \& Haller \cite{Alam06}  have provided an analytical approach to closed-loop separation control, based on a kinematic theory of unsteady separation \cite{Hal04}, most experimental approaches rely  either on low-order reduced models, extracted by physical means  or using identification methods \cite{Juillet13}, or the design of black box controllers \cite{hen07,Gau13a}.

Open-loop control has been successfully applied to separation control: harmonically pulsed synthetic jets 
\cite{Seifert96, Garnier12},
as well as steady suction or blowing at the wall 
(e.g. \cite{McLachlan89} and other references in \cite{Fie90}, or \cite{Wilson13} as a combination of steady suction and pulsed blowing). The determination of the best placement and frequency of the actuators was often left to extensive parameter sweeps resorting to intensive experimental or computational campaigns.

Passive control strategies rely on the optimisation of the geometry or on the addition of appendices, like vortex generators \cite{Pujals10Exp}. 
Their efficient and robust design requires ideally so-called sensitivity maps.
These maps allow one to choose most sensitive regions and therefore to design optimal control configurations, but again to the expense of intensive experimental or computational campaigns. 
A recent example is provided by Parezanovi\'{c} \& Cadot \cite{Par12}, who studied the influence of a thin control wire on the frequency, drag and recirculation length of the wake behind a D-shaped cylinder. 

In all these control methods, an essential aspect is determining meaningful and unambiguous control variables. 
Among those the manipulation of the separation or reattachment locations \cite{Wang03, Alam06}, the recirculation length \cite{Bou14} or the recirculation area \cite{Gau13a} appear as valuable and accessible geometric descriptors of the flow. 
 The existence of a link between the geometric properties of separated flows and their stability properties and associated aerodynamic loads is indeed now well accepted. 

It is known for instance that the destabilisation of the wake of a bluff body beyond the critical Reynolds number takes place simultaneously with a decrease of the recirculation length caused by the mean flow distortion maintained by the progressive development of the instability \cite{Zie97}.
 Therefore, if one is willing to enhance mixing or reduce flow-induced structural vibrations, then it is natural to 
target the recirculation length. 
More recently, Parezanovi\'{c} \& Cadot  \cite{Par12} established a clear correlation between base pressure increase (and therefore drag reduction) and mean recirculation area increase in the wake of a D-shaped cylinder at $\Rey \sim 10^5$, suggesting the direct targeting of the separation properties as a promising control strategy. 
It is also worth noticing that, in some control attempts \cite{choi99,Pas13}, without being directly targeted, a modification of the recirculation length was observed as a by-product of the control scheme.

The present study is dedicated to the determination  of analytical expressions for the sensitivity to steady actuation of the following six geometric indicators of flow separation: the locations of the two separation points which connect the separatrix, the separation angles prevailing at these separation points, the backflow area and the recirculation area. 
These geometric quantities can easily be measured in experiments.
Sensitivity maps are computed by solving adjoint equations at the same cost as that of computing the uncontrolled flow, and these maps allow one to  identify sensitive regions and predict the effect of small-amplitude control without actually computing any controlled flow.
The main focus of this study is on the rigorous derivation of sensitivity expressions  for steady control, which constitutes a first step towards more general unsteady control.
The flow configuration considered is the boundary layer flow above a wall-mounted bump, on which several open-loop \cite{Bou13} and closed-loop \cite{Pas13} control strategies have been tested numerically.

This paper is organised as follows: geometric quantities of interest are introduced in section~\ref{sec:quantities}; the concept of sensitivity analysis is presented in section~\ref{sec:sens_control}; expressions of sensitivity to base flow modification are derived in section~\ref{sec:sensBF}; results and validation for volume control and wall control are given in section~\ref{sec:results}.

\section{Characteristic quantities in separated flows \label{sec:quantities} }

The evolution of the recirculation length in separated flows with the increase in Reynolds number $\Rey$ is well documented \cite{Tan56, Acr68, nis78, Zie97, bar02, Mar03, Gia07, Pas12}. 
For instance, in the flow around a circular cylinder, the recirculation length is known to increase with $\Rey$ in the steady laminar regime while it starts to  decrease in mean value as $\Rey$ is further increased in the unsteady laminar regime.
In the present study, we turn our attention to several characteristic quantities which describe the separation, in complement to the recirculation length: the locations of the two separation points ($x_s$ and $x_r$) which connect the separatrix, the separation angles ($\alpha_s$ and $\alpha_r$) prevailing at these separation points, the backflow area $A_{back}$ and the recirculation area $A_{rec}$. 

As an archetypical flow configuration, we consider the boundary layer  flow  above a wall-mounted bump studied through DNS \cite{Mar03} and through global stability analysis \cite{Ehr05}.

\begin{figure*}
  \psfrag{y}[r][][1][-90]{$y$}	
  \psfrag{x}[][]{$x$}  \centerline{
	\begin{overpic}[width=13 cm,tics=10]{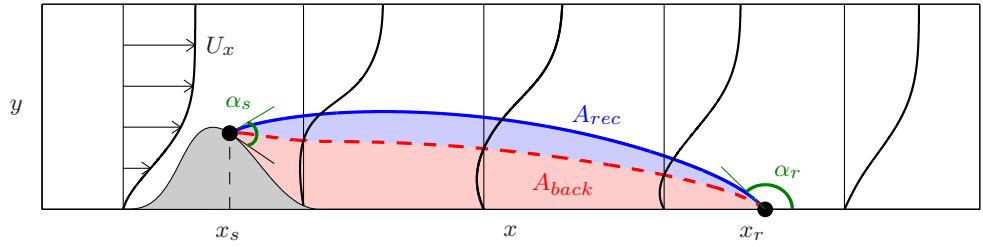}   	
   	\put(21,  20)   {$U_x$}
   	\put(22,   1.2) {$x_s$}
   	\put(75,   1.2) {$x_r$}
   	\put(23,  14.5) {\textcolor[rgb]{0,0.5,0}{$\alpha_s$}}
   	\put(78.5, 7.5) {\textcolor[rgb]{0,0.5,0}{$\alpha_r$}} 
   	\put(58,  13)   {\textcolor{blue}{$A_{rec}$}}   	
   	\put(54,   6.)  {\textcolor{red} {$A_{back}$ }} 
  	\end{overpic}
  }
 \caption{
 Sketch of the bump flow.
The flow separates at $\xx_s$ and reattaches at $\xx_r$ (black dots).
The separatrix (blue solid line) encloses the recirculation region (blue shade) of area $A_{rec}$ and makes angles $\alpha_s$, $\alpha_r$ with the wall.
Velocity profiles are shown with black lines.
The curve where $U_x=0$ (red dashed line) delimits the backflow area (red shade) of area $A_{back}$.
The bump wall geometry is parametrised by $y_w(x)$ and the separatrix by $y_{sep}(x)$.
The axes are not to scale.}
   \label{fig:sketch}
\end{figure*}

\begin{figure}
  \centerline{
   	\begin{overpic}[width=6 cm,tics=10]{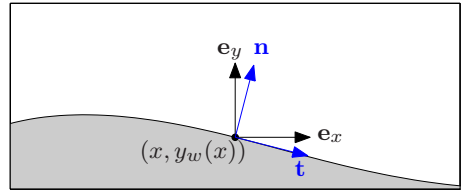}
   	\put(63,3) {\textcolor{blue}{$\tt$}} 	
   	\put(68,11) {$\ex$} 	
   	\put(54,30) {\textcolor{blue}{$\nn$}} 	
    \put(46,30) {$\ey$} 	   	   	
   	\put(29,7) {$(x,y_w(x))$} 	   	
  	\end{overpic}
  }
 \caption{Sketch of tangent and normal vectors to the wall at $(x,y)=(x,y_w(x))$.}
   \label{fig:tn}
\end{figure}

Figure \ref{fig:sketch} is a schematic of typical flow separation. 
The recirculation region is delimited by the wall and the separating streamline. 
This particular streamline, or \textit{separatrix}, makes angles $\alpha_s$ and $\alpha_r$ at the separation point $\xx_{s}$ and reattachment point $\xx_{r}$, respectively.
The wall geometry is described by $y_w(x)$ and the separatrix by $y_{sep}(x)$. 
Unit vectors tangent and normal to the wall are noted $\tt$ and $\nn$ (figure~\ref{fig:tn}), and $\partial_t$, $\partial_n$ stand for  derivatives along $\tt$  and $\nn$.

In this paper we focus on the following quantities:
\begin{enumerate}
\item 
The location of stagnation points, i.e. 
separation   point $\xx_{s}$ and 
reattachment point $\xx_{r}$, 
characterised by zero wall shear stress
\be 
\tau_{s/r} = \partial_n U_t(\xx_{s/r}) = 0;
\label{eq:defxsxr}
\ee
\item 
The angle between the separatrix and the wall at the  separation and reattachment points, given by Lighthill's formula \cite{Lig63}
\be
\tan(\alpha_{s/r})  = 
\left.  
   -3 \frac{ \partial_{nt} U_t } {\partial_{nn} U_t}
\right|_{\xx_{s/r}};
\label{eq:defangle}
\ee
\item 
The area of the backflow region 
\be
\displaystyle A_{back}=  \iint_{\Omega}  \mathbbm{1}_{\Omega_{back}} (\xx) \,\mathrm{d}\Omega
\label{eq:defAb}
\ee
where $\mathbbm{1}_{\Omega_{back}}$ is the characteristic function of $\Omega_{back}=\{(x,y)\,|\,U_x<0\}$;
\item 
The area of the recirculation region enclosed between the separatrix and the wall 
\be 
\displaystyle A_{rec} 
= \iint_{\Omega}  \mathbbm{1}_{\Omega_{rec}}  (\xx) \,\mathrm{d}\Omega
\label{eq:defAr}
\ee
where $\mathbbm{1}_{\Omega_{rec}}$ is the characteristic function of $\Omega_{rec}=\{(x,y)\,|\, x_s\leq x \leq x_r, y_w(x)\leq y \leq y_{sep}(x) \}$.
\end{enumerate}

The steady-state flow $\UU(\xx)$ is calculated by solving the Navier$-$Stokes equations with a finite element method and an iterative Newton procedure on a mesh highly refined near stagnation points.
A two-dimensional  triangulation of the computational domain
is generated with the finite element software \textit{FreeFem++} (http://www.freefem.org),
and equations  are solved in their variational formulation, with the 
following boundary conditions:
Blasius profile at the inlet,
no-slip condition  on the wall,
symmetry condition at the top border,
and convective condition $-P\nn+\nnu\bnabla\UU\nn=\00$ at the outlet.
P2 and P1 Taylor-Hood elements are used for spatial discretisation of  velocity and pressure, respectively
(see also details and validation in \cite{Bou13} and \cite{Bou14}).
Stagnation points are found according to (\ref{eq:defxsxr}) with a bisection on the wall shear stress $\partial_n U_t$.
Angles  are calculated using Lighthill's formula (\ref{eq:defangle}), and are found to agree very well with geometric angles measured between the wall and the separatrix integrated from $\UU(\xx)$ with a fourth-order Runge-Kutta method.
Areas (\ref{eq:defAb})-(\ref{eq:defAr}) are computed with a trapezoidal rule for the two-dimensional integration of the backflow region and recirculation region.

\begin{figure}
  \centerline{
  	\psfrag{y}[r][][1][-90]{$y$}	
	\psfrag{x}[t][]{$x$}
   	\begin{overpic}[width=14. cm,tics=10]{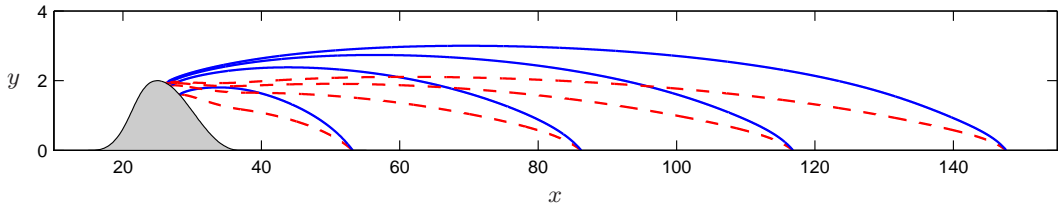}
   	\end{overpic}
   	}
 \caption{Separatrix (blue solid line) and curve of zero streamwise velocity $U_x=0$ (red dashed line), at $\Rey=100, 300, 500$ and $700$. The axes are not to scale.}
   \label{fig:separatrices_re_100_700}
\end{figure}

\begin{figure}
  \centerline{
  	\psfrag{y}[r][][1][-90]{$y$}	
	\psfrag{Re}[t][]{$\Rey$}   
  	\psfrag{as}[r][][1][-90]{$[^{\circ}]$}	
	\psfrag{ar}[r][][1][-90]{$[^{\circ}]$}	
	\psfrag{deg}[r][][1][-90]{$[^{\circ}]$}	
	\psfrag{xs}[r][][1][-90]{ }
	\psfrag{xr}[r][][1][-90]{ }
	\psfrag{ymax}[r][][1][-90]{\textcolor{blue}{ }}
   	\begin{overpic}[width=8.2 cm,tics=10]{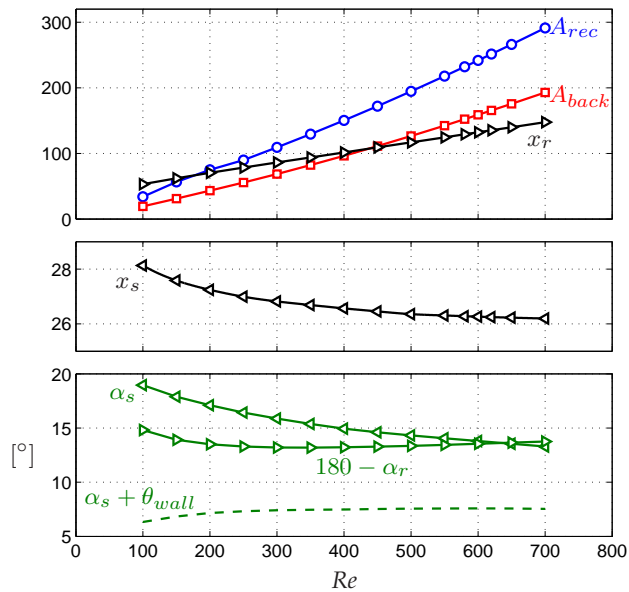}
   		\put(86.5, 89)  {\textcolor{blue}{$A_{rec}$}} 	
	   	\put(86.5, 78.)  {\textcolor{red} {$A_{back}$ }}    	
	   	\put(83, 71)  {$x_r$} 
	   	\put(17, 48){$x_s$} 
   	   	\put(16,30)   {\textcolor[rgb]{0,0.5,0}{$\alpha_s$}} 	
   		\put(49,18)   {\textcolor[rgb]{0,0.5,0}{$180-\alpha_r$}}
	   	\put(12,13) {\textcolor[rgb]{0,0.5,0}{$\alpha_s+\theta_{wall}$}} 	
   	\end{overpic}
  }
 \caption{Variation with Reynolds number of characteristic separation quantities: 
recirculation and backflow areas $A_{rec}$ and $A_{back}$,
separation and reattachment locations $x_s$ and $x_r$,
separation and reattachment angles $\alpha_s$ and $\alpha_r$ relative to the wall (the dashed line shows the 
``absolute'' separation angle $\alpha_s+\theta_{wall}$, i.e. measured relative to $\ex$).
}
   \label{fig:xs_xr_area_vs_re}
\end{figure}

Figure \ref{fig:separatrices_re_100_700}  illustrates  how the backflow and recirculation regions
grow with  Reynolds number in the bump flow.
Figure~\ref{fig:xs_xr_area_vs_re} shows that 
the reattachment point moves downstream linearly with $\Rey$, which is typical of steady separated flows.
The areas $A_{back}$ and $A_{rec}$ show the same trend but increase more than linearly since the backflow and recirculation regions become not only much longer but also slightly higher.
The separation point moves a little upstream but stays  downstream of the bump summit ($x_b=25$).
The reattachment angle is fairly constant, 
$180-\alpha_r \simeq 13-15^{\circ}$.
The separation angle measured relative to the wall decreases from 
$\alpha_s=19^{\circ}$ to $13^{\circ}$ between $\Rey=100$ and 700, but measured relative to $\ex$ it is small and almost constant,  
$\alpha_s+\theta_{wall} \simeq 6-8^{\circ}$. 
It is interesting to note that the separatrix angles vary slowly with $\Rey$ even though the recirculation length increases significantly.
These observations indicate that the main effect of $\Rey$ is to elongate the recirculation region, while the flow remains mostly horizontal.

\section{Sensitivity analysis \label{sec:sens_control}}

In this section, analytical expressions are given for the sensitivity of quantities of interest (\ref{eq:defxsxr})-(\ref{eq:defAr}) to flow modifications, volume forcing and blowing/suction at the wall.

The sensitivity to flow modification of a quantity of interest, say $\phi$,  is a field defined through  the first-order variation   $\delta \phi$ induced by a small flow modification $\bdelta \UU$,
\be
\delta \phi = \left(  \bnabla_\UU \phi \,|\, \bdelta\UU \right),
\label{eq:dphi}
\ee
and it can be computed as:
\be 
\frac{\mathrm{d}\phi}{\mathrm{d}\UU} \bdelta\UU
= 
\lim_{\epsilon\rightarrow 0} 
\frac{ \phi(\UU+\epsilon \bdelta\UU) - \phi(\UU) }{\epsilon}.
\label{eq:angles_sens_BFmod}
\ee
 Here $( \aa \,|\, \bb ) = \iint_{\Omega} \aa \bcdot \bb \,\mathrm{d}\Omega$ 
 denotes   the two-dimensional inner product in $\Omega$.
In other words, the sensitivity 
$\bnabla_\UU \phi = \mathrm{d} \phi/\mathrm{d} \UU$  
is  the gradient of $\phi$ with respect to flow modification. 
Analytical expressions for the sensitivity  to flow modification
of the geometric quantities of interest considered is in this study will be derived in section~\ref{sec:sensBF} (see (\ref{eq:dxs}) for the location of stagnation points, (\ref{eq:alfa_final}) for separatrix angles, and (\ref{eq:sens_Ab})-(\ref{eq:sens_Ar}) for backflow area and recirculation area).

In practice, the base flow cannot be modified arbitrarily and one has to resort to an external control, e.g. passive obstacle,  heating, magnetic field, geometry modification, wall motion, wall blowing or suction, etc. This control in turn alters the velocity field. 
Here we focus on steady control, either in the domain $\Omega$ by means of a body force (source of momentum) $\CC$, or at the wall $\Gamma_w$ by means of blowing/suction with velocity $\UU_c$. However the method is general and easily handles other types of control.

Sensitivities to volume control and wall control can be defined through the variation $\delta\phi$  induced by small-amplitude control,
\be 
	\delta \phi = 	(\bnabla_\CC \phi \,|\, \bdelta \CC) + 
	\langle \bnabla_{\UU_c} \phi \,|\, \bdelta \UU_c \rangle,
	\label{eq:dphi_control}
\ee
where 
  $\langle \aa \,|\, \bb \rangle = \int_{\Gamma_w} \aa \bcdot \bb \,\mathrm{d}\Gamma$ 
  denotes the one-dimensional inner product on $\Gamma_w$.
Sensitivities 
$\bnabla_\CC \phi = \mathrm{d} \phi/\mathrm{d} \CC$ and 
$\bnabla_{\UU_c} \phi = \mathrm{d} \phi/\mathrm{d} \UU_c$ 
can be computed as:
\begin{align} 
& \frac{\mathrm{d}\phi}{\mathrm{d}\CC} \bdelta\CC
= \lim_{\epsilon\rightarrow 0} 
\frac{ \phi(\CC+\epsilon \bdelta\CC) - \phi(\CC) }{\epsilon},
\quad
 \frac{\mathrm{d}\phi}{\mathrm{d}\UU_c} \bdelta\UU_c
= \lim_{\epsilon\rightarrow 0} 
\frac{ \phi(\UU_c+\epsilon \bdelta\UU_c) - \phi(\UU_c) }{\epsilon}.
\end{align} 
Taking into account the definition of $\phi$, and enforcing the  Navier$-$Stokes equations to be satisfied by the flow, a Lagrangian method yields the sensitivities
\begin{align}
\begin{split}
 \bnabla_\CC \phi = \UUa,
\quad
 \bnabla_{\UU_c} \phi = -\Pa \nn - \nnu \bnabla \UUa  \nn,
\label{eq:sens_DF_DUw}
\end{split}
\end{align}
where the  adjoint  flow $(\UUa,\Pa)^T$ is solution of the non-homogeneous  linear 
equations 
\be 
	\bnabla \bcdot \UUa  =0,
 	\quad
  	- \UU  \bcdot \bnabla \UUa    
  	+ \UUa \bcdot \bnabla \UU^T  
	- \bnabla \Pa - \nnu \bnabla^2 \UUa = \bnabla_\UU \phi,
\label{eq:adjBF}
\ee
with  boundary condition
$\UUa=\00$                                               
at the wall.
The forcing term in (\ref{eq:adjBF}) is the sensitivity of $\phi$ to flow modification defined in (\ref{eq:dphi})-(\ref{eq:angles_sens_BFmod}), 
which must therefore be computed beforehand, using expressions derived in detail in section~\ref{sec:sensBF} for all quantities of interest (\ref{eq:defxsxr})-(\ref{eq:defAr}).
Using the same finite element method as for the determination of the base flow, sensitivities  (\ref{eq:sens_DF_DUw}) are obtained by solving the adjoint equations (\ref{eq:adjBF}) in weak form, particularly convenient to express the forcing term $\bnabla_\UU \phi$.

\section{Derivation of sensitivities to flow modification \label{sec:sensBF} }

In this section, analytical expressions are derived for the sensitivity 
of quantities of interest  (\ref{eq:defxsxr})-(\ref{eq:defAr}) to flow modification.
Recall that the wall $\Gamma_w$ is parametrised by $(x,y)=(x,y_w(x))$, as shown in figure \ref{fig:tn}.

\subsection{Stagnation points \label{sec:xsxr}}

\begin{figure}
  \centerline{
	\psfrag{as [deg]}[r][][1][-90]{ }
	\psfrag{tau}[][][1][-90]{}
	\psfrag{x} [][]{$x$}
   	\begin{overpic}[width=8 cm,tics=10]{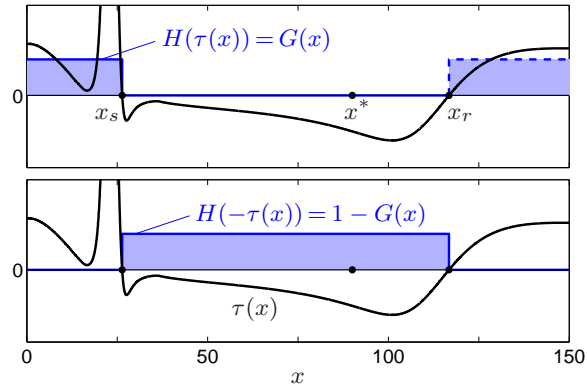}
   	\put(29.5,57)   {\textcolor{blue}{$H(\tau(x))=G(x)$}} 
   	\put(18.5,45)   {$x_s$} 	
   	\put(59.5,45)   {$x^*$}
   	\put(76.5,45)   {$x_r$} 
   	\put(41,  12.5) {$\tau(x)$}   	
   	\put(35,  28)   {\textcolor{blue}{$H(-\tau(x))=1-G(x)$}} 
  	\end{overpic}
  }
 \caption{Wall shear stress and associated Heaviside functions appearing in the expression of the stagnation points (\ref{eq:xs})-(\ref{eq:xr}). }
 \label{fig:HG}
\end{figure}

As expressed in (\ref{eq:defxsxr}), steady separation and reattachment points 
$\xx_{s} = ( x_{s}, y_w(x_{s}) )$ and
$\xx_{r} = ( x_{r}, y_w(x_{r}) )$
are characterised by zero wall shear stress.
Following \cite{Bou14}, stagnation points are redefined in terms of characteristic functions:
\begin{align}	
x_{s} &= \int_{0}^{x^*} H     \left( \tau(x) \right) \,\mathrm{d}x 
         	 =  \int_{0}^{x^*} G \left(       x  \right) \,\mathrm{d}x,  
         	  \label{eq:xs}  
         	 \\
	x_{r} &= \int_{x^*}^{\infty} H   \left( - \tau(x) \right) \,\mathrm{d}x + x^*
         	 =  \int_{x^*}^{\infty} 1-G \left(         x  \right) \,\mathrm{d}x + x^*, 
         	 \label{eq:xr}
\end{align}
where $ \tau(x) = \left. \partial_n U_t \right|_{(x,y_w(x))}	$
is the wall shear stress,
$H$ is the Heaviside step function defined as 
$H(\dumb<0)=0$, 
$H(\dumb>0)=1$,
and $x^*$ is any streamwise location inside the recirculation region.
As illustrated in figure~\ref{fig:HG}, 
the integrand  in (\ref{eq:xs}) is equal to 1 upstream of the separation point, therefore integrating over $0\leq x\leq x^*$  does indeed yield the coordinate $x_s$. 
Similarly, the integrand  in (\ref{eq:xr}) is equal to 1 upstream of the reattachment point, and integrating in $x$ yields the coordinate $x_r$.

We assume for the sake of clarity that reattachment occurs far enough downstream  where the wall is horizontal and 
$\tau(x_r) = \left. \partial_y U_x \right|_{(x_r,y_w(x_r))}$, 
which is verified in practice for all the Reynolds numbers considered.
A flow modification $\bdelta\UU$ makes the  reattachment point move by the following amount:
\begin{align} 
 	\delta  x_{r}
& =  \lim_{\epsilon \rightarrow 0} \frac{ x_{r}(\UU+\epsilon \bdelta \UU)-x_{r}(\UU)}{\epsilon}
\label{eq:dx0}
 \\
& =  \lim_{\epsilon \rightarrow 0} 
\frac{1}{\epsilon} \int \left[ H \left( - \tau(x) -\epsilon \delta\tau(x)  \right) -  H \left( -\tau(x) \right) \right] \,\mathrm{d}x
\label{eq:dx1}
 \\
& =  
\int -\left.\frac{ \mathrm{d}H }{ \mathrm{d}\dumb }\right|_{\dumb=-\tau(x)} \,\delta \tau(x) \,\mathrm{d}x
 \label{eq:dx2}
 \\
& =  
\int  -\left(  \frac{\mathrm{d} \tau}{\mathrm{d} x}(x) \right)^{-1} 
\frac{ \mathrm{d}G }{ \mathrm{d}x }(x) \,\delta \tau(x) \,\mathrm{d}x
\label{eq:dx3} 
\\
& =  
\int  - \left(  \frac{\mathrm{d} \tau}{\mathrm{d} x}(x) \right)^{-1}  \delta(x-x_{r}) \,\delta \tau(x) \,\mathrm{d}x
\label{eq:dx4} 
\\
& =     -   \frac{ \delta \tau(x_{r}) } { \mathrm{d}_x \tau|_{(x_r,y_w(x_r)}}
\label{eq:dx5} 
\end{align}
where (\ref{eq:dx3}) comes from the chain rule differentiation  
$\displaystyle \frac{\mathrm{d} (1-G)}{\mathrm{d} x}(x) 
= -\left.\frac{\mathrm{d} H}{\mathrm{d} \dumb}\right|_{\dumb=-\tau(x)} 
         \frac{\mathrm{d} \tau}{\mathrm{d} x}(x)$,
and (\ref{eq:dx4}) is the result of 
$\displaystyle \frac{\mathrm{d} G}{\mathrm{d} x}(x)=\delta(x-x_{r})$ 
with $\delta(x)$ the Dirac delta function, since $G$ increases from 0 to 1 at $x=x_{r}$.
Finally:
\begin{eqnarray}
 \delta  x_{r} 
= 
( \bnabla_\UU x_{r} \,|\, \bdelta\UU)
=
 -\left.  
 \frac{ \partial_y \delta U_x } { \partial_{xy} U_x} \right|_{\xx_{r}}.
 \label{eq:dxr}
\end{eqnarray}

The variation of the separation point is obtained in a similar way, with only slight sign differences. 
First, the chain rule derivation of $G(x)=H(\tau(x))$   reads $\displaystyle \frac{\mathrm{d} G}{\mathrm{d} x}(x) 
= \left.\frac{\mathrm{d} H}{\mathrm{d} \dumb}\right|_{\dumb=\tau(x)} \frac{\mathrm{d} \tau}{\mathrm{d} x}(x)$.
Second, the expression in terms of Dirac delta is   $\displaystyle \frac{\mathrm{d} G}{\mathrm{d} x}(x)=-\delta(x-x_{s})$ 
since $G$ decreases from 1 to 0 at $x=x_{s}$.
Taking into account the  wall geometry, one obtains:
\begin{eqnarray} 	
 \delta  x_{s}
= 
( \bnabla_\UU x_{s} \,|\, \bdelta\UU)
=
\left. - \frac{1}{\sqrt{1+y_w'^2}} \frac{ \partial_n \delta U_t } {\partial_{nt} U_t} \right|_{\xx_{s}}
\label{eq:dxs}
\end{eqnarray}
where $y_w'=\mathrm{d}y_w/\mathrm{d}x$.
Expression (\ref{eq:dxs}) is valid for the reattachment point too, where $y_w'=0$, 
$\tt\equiv\ex$, and 
$\nn\equiv\ey$.

\begin{figure*}
  \centerline{
  	\psfrag{y}[r][][1][-90]{$y$}
	\psfrag{x}[][]       {$x$} 
  	\begin{overpic}[width=13 cm,tics=10]{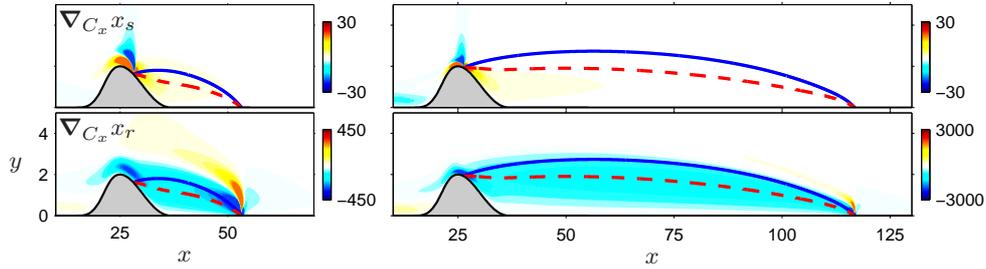}
   		\put(5.5,25)  {$\bnabla_{C_x} x_s$}
   		\put(5.5,14)  {$\bnabla_{C_x} x_r$}
  	\end{overpic}  	
  }
\caption{Sensitivity of stagnation points $x_s$, $x_r$,
  with respect to streamwise volume control $C_x$. 
Left: $\Rey=100$, right: $\Rey=500$.
  The blue solid line is the separatrix, the red dashed line is the curve where $U_x=0$.
}
   \label{fig:DU-xs}
\end{figure*}

Figure~\ref{fig:DU-xs} shows the sensitivity of stagnation points to volume control in the streamwise direction  $C_x=\CC \bcdot \ex$
obtained at $\Rey=100$ and 500 using  (\ref{eq:dxr})-(\ref{eq:dxs}) and the method presented in section~\ref{sec:sens_control}.
Red (resp. blue) regions indicate where a localised, small-amplitude body force oriented along $\ex$ would move stagnation points upstream, $\delta x_{s/r}>0$ (resp. downstream, $\delta x_{s/r}<0$).
The separation point is mostly sensitive near $\xx_s$. 
The reattachment point is sensitive near $\xx_r$,  but also at the bump summit, in the whole shear layer along the separatrix, and in the recirculation region.

\subsection{Separation and reattachment angles \label{sec:angle}}

It is remarkable but not well known that the angle between
the separatrix and the wall can be expressed analytically as a function of flow quantities at the stagnation point, as expressed in (\ref{eq:defangle}).
We recall briefly Lighthill's original presentation \cite{Lig63}.
For the sake of simplicity we  assume first that the wall is flat and horizontal, $y_w(x)=0$. A Taylor expansion of the streamwise velocity near the wall, $y\ll 1$, reads
\begin{equation}
U_x(x,y) = U_x(x,0) +  \partial_y U_x(x,0) \, y +   \partial_{yy} U_x(x,0) \, \frac{y^2}{2} 
+ \mathcal{O}(y^3).
\end{equation}
This expression is conveniently recast using
(i) the no-slip condition $U_x(x,0)=0$,
(ii) the vorticity 
$\omega(x,0)=-\partial_y U_x(x,0)$,
and 
(iii) the streamwise momentum equation 
$\partial_{yy} U_x(x,0) = \Rey\partial_x p(x,0)$:
\begin{eqnarray}
U_x(x,y) =  -\omega(x,0) \, y +  \frac{\Rey}{2} \partial_x p(x,0) \, y^2 + \mathcal{O}(y^3).
\end{eqnarray}
Equivalently, the stream function defined as $U_x=\partial_y \psi$ reads
\be
\psi(x,y) = -\frac{1}{2} \omega(x,0) \,y^2  + \frac{\Rey}{6}\partial_x p(x,0)  \,y^3 
+ \mathcal{O}(y^4).
\ee
The separatrix $\psi=0$ is thus described by 
$y_{sep}(x)=3 \omega(x,0) / \Rey \partial_x p(x,0)$ and
separates from or reattaches to the wall with the angle $\alpha$ such that
\begin{align}
\begin{split}
\tan(\alpha) 
&= \frac{\mathrm{d}y_{sep}}{\mathrm{d}x}
=  \frac{3}{\Rey} \left(
 \frac{\partial_x    \omega }{\partial_x    p} 
-  \omega  \frac{  \partial_{xx}p }{(\partial_x    p)^2}  
 \right)
\\
&=  \frac{3}{\Rey} \frac{\partial_x    \omega }{\partial_x    p}
= -3 \frac{\partial_{xy} U_x    }{\partial_{yy} U_x}
\label{eq:lighthill}
\end{split}
\end{align}
because $\omega(x_{s/r},0)=0$.
Equation (\ref{eq:lighthill}) is valid for a curved or inclined wall too, hence recovering (\ref{eq:defangle}).
A longer but similar derivation is possible following the steps of \cite{Hal04} for unsteady flows, and of course the same expression is also obtained if taking all quantities as steady in his final expression.

To derive the sensitivity of the angle, we introduce the   function $f$ defined on the wall
\be
\displaystyle \left. f(\UU,x) = \atan \left( -3 \frac{ \partial_{nt} U_t } {\partial_{nn} U_t}\right|_{(x,y_w(x))}\right),
\ee
 where  $\UU$ and $x$ are treated as independent  variables.
The separation and reattachment angles are equal to $\alpha_{s/r}=f(\UU,x_{s/r})$.
Their variation with flow modification is 
\begin{eqnarray}
\delta \alpha_{s/r} 
=
(\bnabla_\UU \alpha_{s/r} \,|\, \bdelta \UU)
= \left( 
\frac{\partial f}{\partial \UU} 
+ \left. \left. \frac{\partial f}{\partial x}\right|_{x_{s/r}} 
\frac{\mathrm{d} x_{s/r}}{\mathrm{d} \UU}
\,\right|\, 
\bdelta \UU \right),
\label{eq:D_u_alpha}
\end{eqnarray}
where the first term of the sensitivity is the direct angle variation due to the change in tangential velocity  $U_t$, while the second term is the indirect angle variation due to the displacement of the stagnation points $\xx_{s/r}$.
Before deriving in detail  each of the terms of (\ref{eq:D_u_alpha})  we give their expression below:
\begin{align} 
\left(\left.
\frac{\partial f}{\partial \UU}		
\,\right|\,  \bdelta\UU \right)
&=
\left. -  3 \frac{ A \, \partial_{nt} \delta U_t 
	             -B \, \partial_{nn} \delta U_t 
	            } 
	            {A^2+9B^2} \right|_{\xx_{s/r}},
	            \label{eq:duf}
\\
 \left. \frac{\partial f}{\partial x}\right|_{\xx_{s/r}} \delta x_{s/r} 
&= 
  -3 \frac{   B'A -  A'B }
     { A^2 + 9B^2  } \, \delta x_{s/r},
      \label{eq:dxf}
\\
\delta x_{s/r} 
&= \left(\left. \frac{\dd x_{s/r}}{\dd \UU} \,\right|\, \bdelta\UU \right)
= \left(\left. \bnabla_\UU x_{s/r} \,\right|\, \bdelta\UU \right),
 \label{eq:dux}
\end{align} 
where
\begin{align} 
A(x) &= \partial_{nn} U_t(x,y_w(x)), 
\\
B(x) &= \partial_{nt} U_t(x,y_w(x)),
\\
 A'(x)  &= 3\frac{y_w''}{1+y_w'^2} \partial_{nn} U_n + \sqrt{1+y_w'^2} \partial_{nnt} U_t, 
\\
 B'(x) &= \frac{y_w''}{1+y_w'^2} (\partial_{nn} U_t+2\partial_{nt} U_n) - \sqrt{1+y_w'^2} \partial_{nnt} U_n. 
 \label{eq:dxB}
\end{align}

Combining equations (\ref{eq:D_u_alpha})-(\ref{eq:dxB})  yields:
\begin{align}
\begin{split}
\delta\alpha_{s/r}  
&= \left( \bnabla_\UU \alpha_{s/r}  \,|\, \bdelta\UU \right)
\\ 
&= 
-\frac{3}{A^2 + 9B^2}
\left.
  \left(	
    A\partial_{nt} -B\partial_{nn}
    +\frac{A'B-B'A}{B\sqrt{1+y_w'^2}}     \partial_n
  \right)
  \delta U_t 
\right|_{\xx_{s/r}}.
\end{split}
\label{eq:alfa_final}
\end{align}

\begin{figure*}
  \centerline{
  	\psfrag{y}[r][][1][-90]{$y$}
	\psfrag{x}[][]       {$x$} 
  	\begin{overpic}[width=13 cm,tics=10]{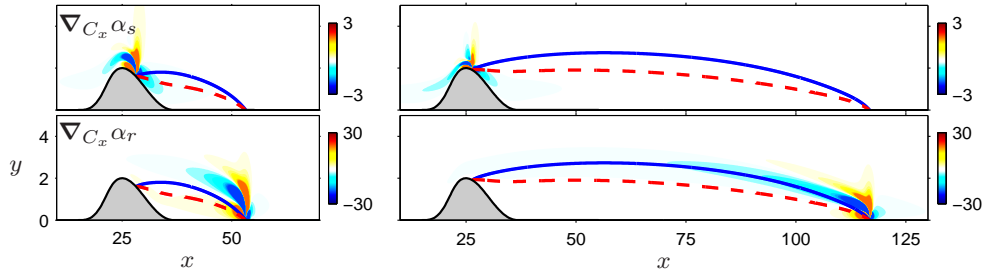}
   		\put(5.5,25)  {$\bnabla_{C_x} \alpha_s$}
   		\put(5.5,14)  {$\bnabla_{C_x} \alpha_r$}
  	\end{overpic}
  }
\caption{Sensitivity of separation and reattachment angles $\alpha_s$, $\alpha_r$,
  with respect to streamwise volume control $C_x$. 
Left: $\Rey=100$, right: $\Rey=500$.
  The blue solid line is the separatrix, the red dashed line is the curve where $U_x=0$.
}
   \label{fig:DU-xalfa}
\end{figure*}

Figure~\ref{fig:DU-xalfa} shows the sensitivity of separatrix angles to streamwise volume control
obtained at $\Rey=100$ and 500 using  (\ref{eq:alfa_final}).
Like stagnation points, the separation angle is mostly sensitive near $\xx_s$, and
the reattachment angle near $\xx_r$. 
These sensitivity maps show complex structures, with regions of opposite signs close to each other. This  indicates that a small displacement of the forcing location could result in changing the sign of the  variation $\delta \alpha$.

We now turn to the derivation of the three terms in (\ref{eq:D_u_alpha}).
The latter term (\ref{eq:dux}) is precisely the variation of stagnation points
(\ref{eq:dxr})-(\ref{eq:dxs}).
Next, the variation of $f$ with flow modification (at fixed $x$) is:
\begin{eqnarray}
\delta f = \left( \left. \frac{\partial f}{\partial \UU}		 \,\right|\, \delta\UU  \right)
		 = \lim_{\epsilon \to 0} \frac{f(\UU+\epsilon \delta\UU,x_{s/r}) - f(\UU,x_{s/r})}{\epsilon}, 	
\end{eqnarray}
where, at first order:
\begin{align}
\begin{split}
&f(\UU+\epsilon \delta\UU, x_{s/r}) 
= \atan \left( \left. -3 \frac{ \partial_{nt} (U_t+\epsilon \delta U_t)} 
{\partial_{nn} (U_t+\epsilon \delta U_t)} \right. \right)
\\
%
	&= \atan \left(-3 \frac{\partial_{nt} U_t} {\partial_{nn} U_t} 
	                -3 \epsilon \frac{ \partial_{nn} U_t \, \partial_{nt} \delta U_t 
	                                  -\partial_{nt} U_t \, \partial_{nn} \delta U_t } 
	                				 {(\partial_{nn} U_t)^2} 	                	
	           \right)
	\\
	\quad 
	&=\displaystyle
	\atan \left( -3 \frac{\partial_{nt} U_t} {\partial_{nn} U_t}  \right)
	    - \frac{3\epsilon}{1+ \left(-3 \frac{\partial_{nt} U_t} {\partial_{nn} U_t} 
	    \right)^2} 
          \frac{ \partial_{nn} U_t \, \partial_{nt} \delta U_t 
	            -\partial_{nt} U_t \, \partial_{nn} \delta U_t 
	           } 
        	   {(\partial_{nn} U_t)^2}
\\
&= \atan \left( -3 \frac{\partial_{nt} U_t} {\partial_{nn} U_t}  \right)
\left. -  3 \epsilon \frac{ \partial_{nn} U_t \, \partial_{nt} \delta U_t 
	             -\partial_{nt} U_t \, \partial_{nn} \delta U_t 
	            } 
	            {(\partial_{nn} U_t)^2 + 9(\partial_{nt} U_t)^2}, 
	            \right.    
\end{split}
\end{align}
which yields expression (\ref{eq:duf}).

Finally, the variation of $ f(\UU,x) = \atan \left( -3  B(x)/A(x) \right)$ with $x$ (for fixed flow conditions) is derived in a similar way, with straightforward composition of derivatives of $\atan$ and of a quotient:
\begin{align} 
\delta f
=
 \left. \frac{\partial f}{\partial x}\right|_{x_{s/r}} \delta x_{s/r} 
&=  
-3\frac{1}{1+(3B/A)^2} \frac{A B' - B  A'}{A^2}
\\
&=  -3 \frac{   B'A -  A'B }
     { A^2 + 9B^2  } \, \delta x_{s/r},
\end{align}
which yields expression (\ref{eq:dxf}).
However, some  care is needed when computing the derivatives of $A$ and $B$.
Although the streamwise derivative of the \textit{total} velocity field at the wall  $\partial_x \UU$ is related in a simple way to the tangential derivative $\partial_t \UU$ by geometric considerations,
\be 
\left. \mathrm{d}_x  \UU \right|_{(x,y_w(x))} 
= (\partial_x+y_w'\partial_y) \UU 
= \sqrt{1+y_w'^2} \,\partial_t \UU,
\ee
this is  true 
neither for \textit{individual} velocity components 
nor     for velocity \textit{derivatives}. 
For example:
\be 
\left. \mathrm{d}_x  (\partial_{n} U_t) \right|_{(x,y_w(x))} 
\neq 
(\partial_x+y_w'\partial_y) (\partial_{n} U_t) 
= \sqrt{1+y_w'^2} \, \partial_t (\partial_{n} U_t).
\label{eq:incorrect_deriv}
\ee
This is because  the tangential velocity 
$U_t = \UU \bcdot \tt$
depends on $x$ not only through $\UU$ but also through the local tangent vector $\tt=\tt(x)=\tt(x,y_w(x))$.
Similarly, the normal derivative 
$\partial_n = \bnabla\bcdot\nn$ 
depends on $x$ because the normal vector $\nn=\nn(x)=\nn(x,y_w(x))$ does.
In (\ref{eq:incorrect_deriv}), one must therefore take into account $\mathrm{d}_x \tt$ and $\mathrm{d}_x \partial_n$. 
The calculation is straightforward but tedious when expressing all quantities in the basis $(\ex,\ey)$; instead, one can make a systematic use of the nabla operator:
\begin{align}
\begin{split}
\mathrm{d}_x (\partial_n U_t) 
={}& \mathrm{d}_x \left( (\bnabla\bcdot\nn) (\UU\bcdot \tt) \right)
\\
={}& \bnabla\bcdot (\mathrm{d}_x \nn) (\UU\bcdot \tt) 
+ (\bnabla\bcdot\nn) (\mathrm{d}_x \UU\bcdot \tt) 
+ (\bnabla\bcdot\nn) ( \UU\bcdot \mathrm{d}_x\tt) 
\\
={}& \bnabla\bcdot \left(-\frac{y_w''}{1+y_w'^2}\tt \right) (\UU\bcdot \tt) 
+ (\bnabla\bcdot\nn) \left(\sqrt{1+y_w'^2} \,\partial_t \UU\bcdot \tt \right)  
+ (\bnabla\bcdot\nn) \left( \UU\bcdot \frac{y_w''}{1+y_w'^2}\nn \right) 
\\
={}&  -\frac{y_w''}{1+y_w'^2}\partial_t  U_t 
+  \sqrt{1+y_w'^2} \,\partial_{nt} U_t
+ \frac{y_w''}{1+y_w'^2} \partial_n  U_n
\\
={}&  \frac{y_w''}{1+y_w'^2} (\partial_n  U_n-\partial_t  U_t )
+  \sqrt{1+y_w'^2} \,\partial_{nt} U_t.
\end{split}
\end{align} 
Compared to (\ref{eq:incorrect_deriv}), two additional  terms coming from the derivatives of $\partial_n$ and $\tt$ clearly appear.
The calculation of $A'$ and $B'$ follows similar steps.

\subsection{Backflow area \label{sec:Ab}}

The backflow area (\ref{eq:defAb}) can be expressed as
\be 
A_{back}	
=  \iint_{\Omega}  \mathbbm{1}_{\Omega_{back}}(\xx) 
\,\mathrm{d}\Omega
= \iint_{\Omega}  H(-U_x(\xx)) 
\,\mathrm{d}\Omega.
\ee
Its sensitivity is derived in the same vein as that of stagnation points (section~4\ref{sec:xsxr}):
\begin{align} 
\begin{split}
\label{eq:dA}
 	\delta A_{back}	
& =  \lim_{\epsilon \rightarrow 0} \frac{A_{back}	(\UU+\epsilon \bdelta \UU)-A_{back}(\UU)}{\epsilon}
 \\
& =  \lim_{\epsilon \rightarrow 0} 
\frac{1}{\epsilon} 
\iint_{\Omega} \left[ H(-U_x(\xx) -\epsilon\delta U_x(\xx)) - H(-U_x(\xx)) \right]
\,\mathrm{d}\Omega
 \\
& =  
\iint_{\Omega}  -\left.\frac{ \mathrm{d}H }{ \mathrm{d}u }\right|_{u=-U_x} \,\delta U_x(\xx) \,\mathrm{d}\Omega
 \\
& =  
\iint_{\Omega} \left(\partial_n U_x \right)^{-1} 
(\bnabla \mathbbm{1}_{\Omega_{back}}(\xx) \bcdot \nn) \,\delta U_x(\xx) \,\mathrm{d}\Omega
\\
& =  
\iint_{\Omega}  - \left( \partial_n U_x \right)^{-1}  
\delta_{\Gamma_{back}}(\xx) \,\delta U_x(\xx) \,\mathrm{d}\Omega
\\
&= -\oint_{\Gamma_{back}} \frac{ \delta U_x}
{\partial_n U_x }  \mathrm{d} \Gamma,
\end{split}
\end{align}
where $\Gamma_{back}$ is the boundary of the backflow region $\Omega_{back}$,
and  $\partial_{n} U_x$ is the outward derivative normal to $\Gamma_{back}$.
Here we used the chain-rule derivation
$\displaystyle 
 \bnabla \mathbbm{1}_{\Omega_{back}}(\xx) \bcdot \nn
=
- \left.\frac{ \mathrm{d}H }{ \mathrm{d}u }\right|_{u=-U_x} 
\partial_n U_x(\xx)$,
and a higher-order generalisation 
of the one-dimensional relation 
\be 
\displaystyle \int\phi(x) 
\frac{ \mathrm{d}H(x-x_0) }{ \mathrm{d}x }  
\,\mathrm{d}x
=\int\phi(x) \delta(x-x_0) \,\mathrm{d}x
=\phi(x_0), 
\ee
namely:
\be 
-\iint_{\Omega} \phi(\xx) \bnabla \mathbbm{1}_{\Omega_{back}}(\xx) \bcdot \nn \,\mathrm{d} \Omega
=
\iint_{\Omega} \phi(\xx) \, \delta_{\Gamma_{back}}(\xx) \,\mathrm{d} \Omega
=
\oint_{\Gamma_{back}} \phi(\xx)
\ee
 where  $\delta_{\Gamma_{back}}$ is the two-dimensional delta function associated to 
$\mathbbm{1}_{\Omega_{back}}$,
 and $\nn$ the outward normal  of $\Gamma_{back}$.
Therefore
\be 
\delta A_{back}	
= \left( \bnabla_\UU A_{back} \,|\, \bdelta\UU \right)
= - \oint_{\Gamma_{back}}
\frac{\delta U_x}{\partial_{n} U_x} 
\,\mathrm{d}\Gamma.
\label{eq:sens_Ab}
\ee

\begin{figure*}
  \centerline{
  	\psfrag{y}[r][][1][-90]{$y$}
	\psfrag{x}[][]       {$x$} 
	\begin{overpic}[width=13 cm,tics=10]{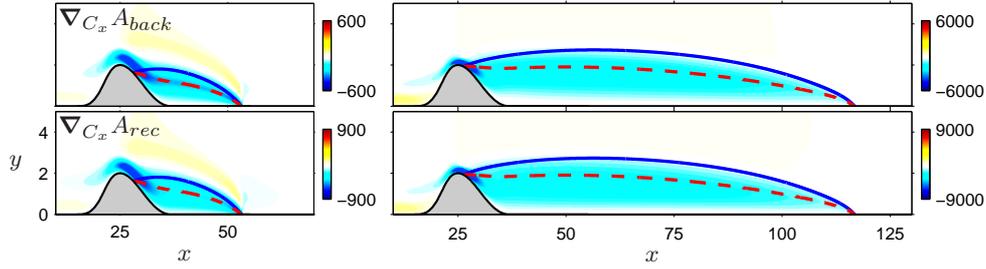}
   		\put(5.5,25)  {$\bnabla_{C_x} A_{back}$}
   		\put(5.5,14)  {$\bnabla_{C_x} A_{rec}$}
  	\end{overpic}
  }
\caption{Sensitivity of backflow and recirculation areas $A_{back}$, $A_{rec}$,
  with respect to streamwise volume control $C_x$. 
Left: $\Rey=100$, right: $\Rey=500$.
  The blue solid line is the separatrix, the red dashed line is the curve where $U_x=0$.
}
   \label{fig:DU-AbAr}
\end{figure*}

The sensitivity of $A_{back}$ to volume control
obtained using  (\ref{eq:sens_Ab}) is shown in figure~\ref{fig:DU-AbAr}.
Regions of large sensitivity extend from upstream  of the bump summit all the way to the reattachment point, with opposite signs below and above the separatrix.

\subsection{Recirculation area \label{sec:Ar} }

We first rewrite the recirculation area (\ref{eq:defAr}) as
\be 
A_{rec} 
= \iint_{\Omega}  \mathbbm{1}_{\Omega_{rec}}(\xx) \,\mathrm{d}\Omega
= \int_{x_s}^{x_r} \int_{y_w(x)}^{y_{sep}(x)} 
\dd y \, \dd x
\ee
where we recall that $y_w(x)$ describes
the wall height and  $y_{sep}(x)$  the separatrix height.
Then we notice that it is possible to give an Eulerian characterisation of the separatrix, namely that the flow rate through any vertical cross section of the recirculation region is zero:
\be
\int_{x_s}^{x_r} 
\int_{y_w(x)}^{y_{sep}(x)} 
U_x(\xx) \, \dd y \, \dd x = 0.
\label{eq:sep_flow_rate}
\ee
The sensitivity of the recirculation area with respect to flow modification is 
\begin{align}
\delta A_{rec}	
& = \lim_{\epsilon \rightarrow 0} \frac{A_{rec}	(\UU+\epsilon \bdelta \UU)-A_{rec}(\UU)}{\epsilon}
 \\
& = \lim_{\epsilon \rightarrow 0} 
\frac{1}{\epsilon} 
\int_{x_s+\epsilon\delta x_s}^{x_r+\epsilon\delta x_r} \int_{y_w(x)}^{y_{sep}(x)+\epsilon\delta y_{sep}(x)} 
\dd y \, \dd x.
\label{eq:sens_Ar_1}
\end{align}
Next,  we use (\ref{eq:sep_flow_rate}) to obtain the first-order variation of the separatrix height:
\be
\delta y_{sep}(x) =  -\frac{1}{U_x(x,y_{sep}(x))} \int_{y_w(x)}^{y_{sep}(x)} \delta U_x(\xx) \, \mathrm{d}y.
\ee
Substituting into (\ref{eq:sens_Ar_1}), 
splitting integration intervals into 
$[y_w, y_{sep}] \cup [y_{sep}, y_{sep}+\epsilon\delta y_{sep}]$
and
$[x_s+\epsilon\delta x_s, x_s] \cup [x_s, x_r] \cup [x_r, x_r+\epsilon\delta x_r]$,
and keeping  first-order terms finally leads to
\begin{equation}
\delta A_{rec}	
= \left( \bnabla_\UU A_{rec} \,|\, \bdelta\UU  \right)
= \int_{x_s}^{x_r} \frac{-1}{U_x(x,y_{sep}(x))} 
\left(\int_{y_w(x)}^{y_{sep}(x)} \delta U_x(\xx) \, \dd y 
\right) \dd x.
\label{eq:sens_Ar}
\end{equation}

The sensitivity of $A_{rec}$ to volume control obtained using  (\ref{eq:sens_Ar}) is shown in figure~\ref{fig:DU-AbAr}.
As could have been expected, it is very similar to the sensitivity of the backflow area.

\section{Results \label{sec:results}}

\subsection{Sensitivity maps}

\begin{figure*}
  \centerline{
  	\psfrag{y}[r][][1][-90]{$y$}
	\psfrag{x}[][]       {$x$} 
	\begin{overpic}[height=9.5 cm,tics=10]{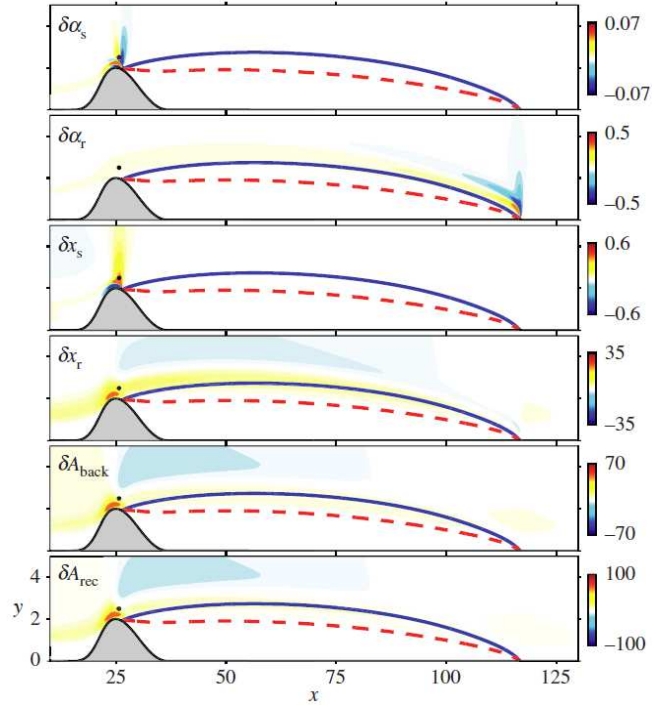}
  	\end{overpic}
  }
\caption{Effect of a control cylinder of diameter $d=0.05$ on separation and reattachment angles $\alpha_s$, $\alpha_r$,
  stagnation points  $x_s$, $x_r$, backflow area $A_{back}$ and recirculation area $A_{rec}$. 
 $\Rey=500$.
  The blue solid line is the separatrix, the red dashed line is the curve where $U_x=0$.
  The black  dot shows the position of the control cylinder
  $(x_c,y_c)=(25.7,2.5)$ used for validation in section 5\ref{sec:valid}.  
}
   \label{fig:SmallCyl}
\end{figure*}

Figures~\ref{fig:DU-xs}, \ref{fig:DU-xalfa} and \ref{fig:DU-AbAr} already presented the sensitivity of all quantities of interest (\ref{eq:defxsxr})-(\ref{eq:defAr}) to volume control.
This sensitivity information can be used to compute the effect of a small control cylinder of diameter $d$ inserted in the flow at $(x_c,y_c)$.
This effect is modelled as a steady volume force opposed to the hypothetical drag force the control cylinder would feel if it were invested by the uniform flow $\UU(x_c,y_c)$:
\be
	\bdelta \CC(x,y) = -\frac{1}{2} d \, C_d(x,y) \, ||\UU(x,y)|| \, \UU(x,y) \, \delta(x-x_c,y-y_c) 
	\label{eq:SmallCylForce}
\ee
where $C_d$ is the cylinder drag coefficient.
Its value depends on the local Reynolds number 
$\Rey_d(x,y)=||\UU_b(x,y)|| \,d/\nu$,
which we compute from a fit of experimental and numerical data \cite{Bou13,Bou14}.
From (\ref{eq:dphi_control}) and (\ref{eq:SmallCylForce}), quantities of interest  vary according to:
\begin{align}
\begin{split}
\delta \phi &= (\bnabla_\CC \phi \,|\, \bdelta \CC)
= -\frac{1}{2} d  \, C_d(\xx_c) \, ||\UU(\xx_c)||  \,
 \bnabla_\CC \phi(\xx_c) \bcdot \UU(\xx_c) .
	\label{eq:SmallCylEffectRL}
\end{split}
\end{align}
Note that  the predicted value $\delta \phi$  varies linearly with the force, by construction, but non-linearly with the diameter of the control cylinder.
Figure~\ref{fig:SmallCyl} shows the effect of a cylinder of diameter $d=0.05$ at $\Rey=500$.
The separation point and separation angle are mostly affected if the cylinder is inserted close to $\xx_s$, and  hardly vary otherwise.
The reattachment angle is sensitive close to $\xx_r$, and is weakly  increased if the control cylinder is located in the shear layer.
 Overall, these three quantities appear robust since they cannot be modified easily (scales next to color bars confirm that their variations are of small amplitude).
The reattachment point is much more sensitive and is predicted to move downstream if the control cylinder is inserted in the shear layer (particularly at the bump summit) or upstream,
and should instead move slightly upstream for a  cylinder farther away from the wall.
Backflow and recirculation areas are affected in a fairly similar way,
increasing when the control cylinder is located near the bump summit or upstream,
and decreasing when the cylinder is farther above the bump or the early recirculation region.

\begin{figure}
  \psfrag{x}[t][b]{$x$}
  \centerline{	
  	\begin{overpic}[width=9 cm,tics=10]{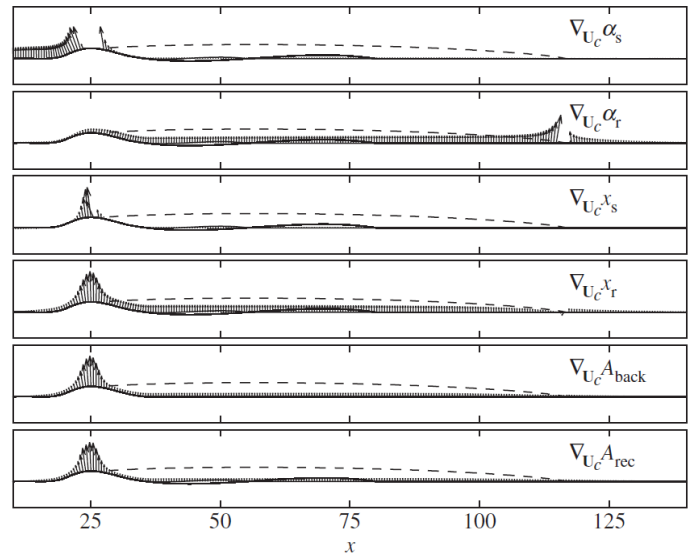}
 	\end{overpic}
  }
 \caption{Sensitivity of separation and reattachment angles $\alpha_s, \alpha_r$, stagnation points $x_s, x_r$, backflow area $A_{back}$ and recirculation area $A_{rec}$ to wall control $\UU_c$.  Arrows point in the direction of positive sensitivity.   The dashed line is the separatrix.  $\Rey=500$.}
   \label{fig:D_Uw-re500-arrows}
\end{figure}

Figure~\ref{fig:D_Uw-re500-arrows} shows sensitivity to wall control.
Arrows  point in the direction of positive sensitivity.
All quantities are significantly more sensitive to normal actuation than to tangential actuation (axes are to scale, so that arrows show the actual orientation relative to the wall).
Separation and reattachment angles are naturally most sensitive close to $\xx_s$ and $\xx_r$, respectively.
More interestingly, $\alpha_s$ is also sensitive upstream of the bump and $\alpha_r$ at the bump summit and in the whole recirculation region.
The separation point $x_s$ is sensitive only at the bump summit.
Finally, $x_r$, $A_{back}$ and $A_{rec}$ are efficiently controlled by wall actuation at the bump summit and to a lesser extent in the whole recirculation region; unsurprisingly, these three quantities have very similar sensitivities.
Note that the sensitivities of $\alpha_s$ and $x_s$ are very large at $x_s$, and the sensitivity of $\alpha_r$ is very large at $x_r$;  for clarity, longest arrows at these locations are not shown.

Figures~\ref{fig:SmallCyl}-\ref{fig:D_Uw-re500-arrows} allow to identify regions where quantities of interest are  affected the most by control and to which extent.

\subsection{Validation and control \label{sec:valid} }

In this section, we illustrate how control configurations can be designed based on sensitivity information.
We also  validate the method by comparing sensitivity predictions against non-linear results obtained from actually controlled flows at $\Rey=500$.

\begin{figure}
  \centerline{
	\psfrag{deg}[r][][1][-90]{$[^{\circ}]$}
	\psfrag{xr}[r][][1][-90]{ }
	\psfrag{xs}[r][][1][-90]{ }	
	\psfrag{W} [t][b]{flow rate $W$}
   	\begin{overpic}[width=8 cm,tics=10]{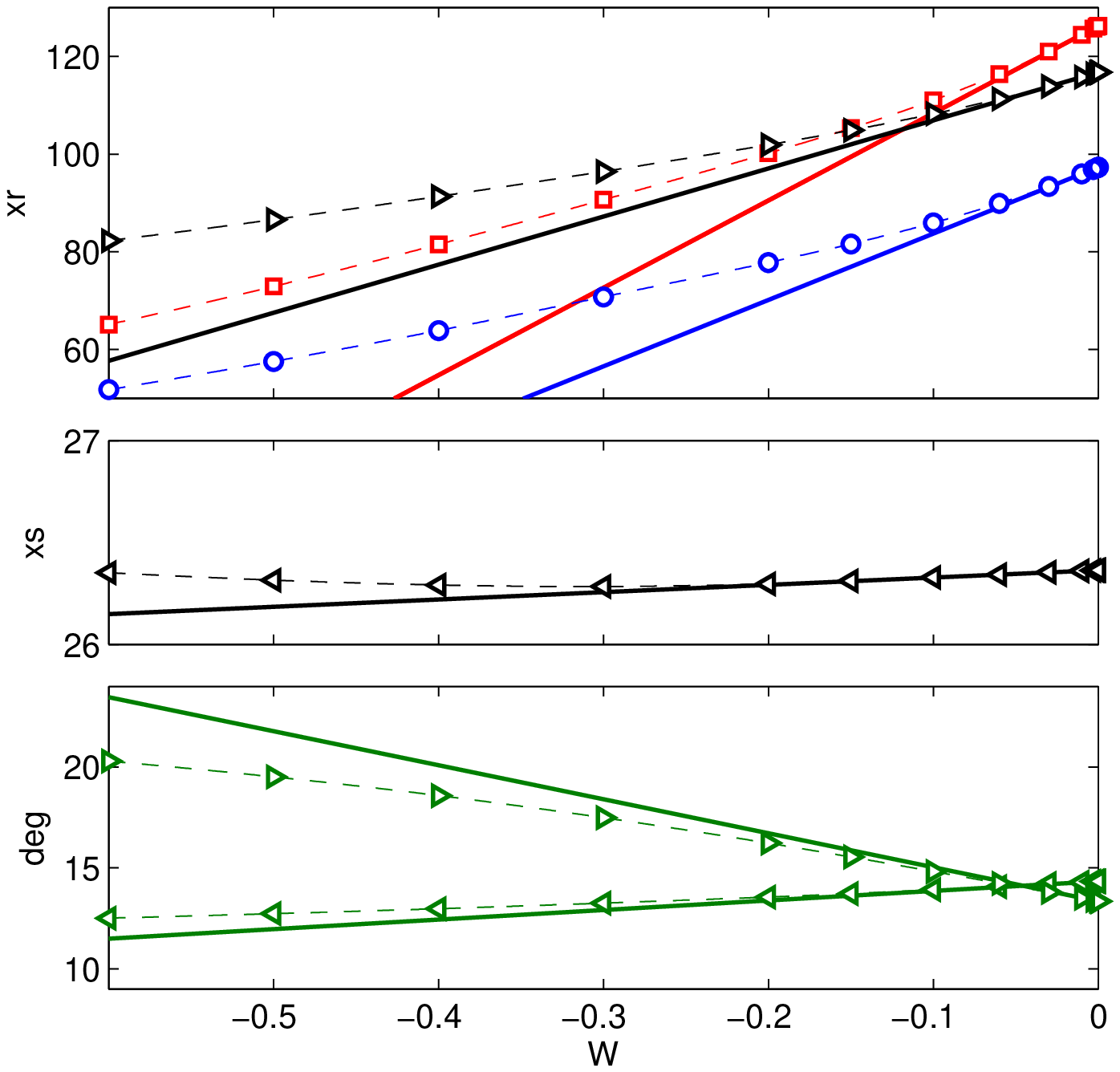}
	\put(-5, 91)  {$(a)$} 
   	\put(99, 92)  {\textcolor{red} {$A_{back}$ }} 
   	\put(99, 79)  {\textcolor{blue}{$A_{rec}/2$}} 	   	   	
   	\put(99, 87)  {$x_r$} 
   	\put(15, 46)  {$x_s$} 
   	\put(58, 25.) {\textcolor[rgb]{0,0.5,0}{$180-\alpha_r$}}
   	\put(16, 16)  {\textcolor[rgb]{0,0.5,0}{$\alpha_s$}} 	
  	\end{overpic}
  }
  \vspace{0.5cm}
  \centerline{
	\psfrag{y}[r][][1][-90]{$y$}	
	\psfrag{x} [t][b]{$x$}
   	\begin{overpic}[width=9.5 cm,tics=10]{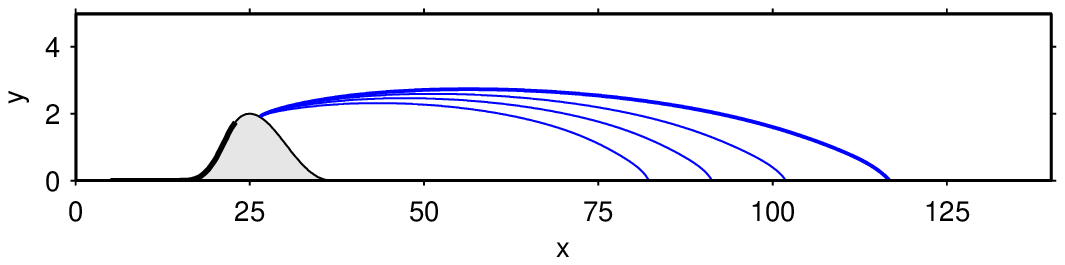}
   	\put(-5, 22)  {$(b)$} 
  	\end{overpic}
  }
  \caption{Effect of  wall suction  applied over $5 \leq x \leq 23$ with total flow rate $W$, at $\Rey=500$.
  $(a)$
Variation of characteristic separation quantities (recirculation and backflow areas, stagnation locations, separation and reattachment angles).
  Thick solid lines show theoretical predictions from sensitivity analysis, while symbols show results from non-linear calculations.
   $(b)$
     Separatrix  for the uncontrolled flow ($W=0$, thick line), and $W=-0.2$, -0.4 and -0.6 (thin lines).
   }
   \label{fig:wall_control}
\end{figure}

Figure~\ref{fig:wall_control} shows how separatrix angles, 
stagnation locations, and backflow and recirculation areas vary when applying small-amplitude vertical wall suction $(U_c=0, V_c<0)$ over $5 \leq x \leq 23$, with total flow rate $W$.
All quantities decrease, although not by the same amount: the reattachment point moves significantly upstream, inducing a large reduction in backflow and recirculation areas.
Separatrix angles decrease only slightly.
The separation point is virtually fixed, reminiscent of  the fact that it is fairly independent of  $\Rey$ in the uncontrolled case (figure~\ref{fig:xs_xr_area_vs_re}).
The agreement between sensitivity predictions (straight solid lines) and actual results (symbols) is excellent at small flow rate. However, non-linear effects are non-negligible when $|W| \gtrsim 0.1-0.2$ and in all cases, make actual variations smaller than predicted by sensitivity analysis.

\begin{figure}
  \centerline{
	\psfrag{deg}[r][][1][-90]{$[^{\circ}]$}
	\psfrag{xr}[r][][1][-90]{ }
	\psfrag{xs}[r][][1][-90]{ }	
	\psfrag{d} [t][b]{cylinder diameter $d$}
   	\begin{overpic}[width=8. cm,tics=10]{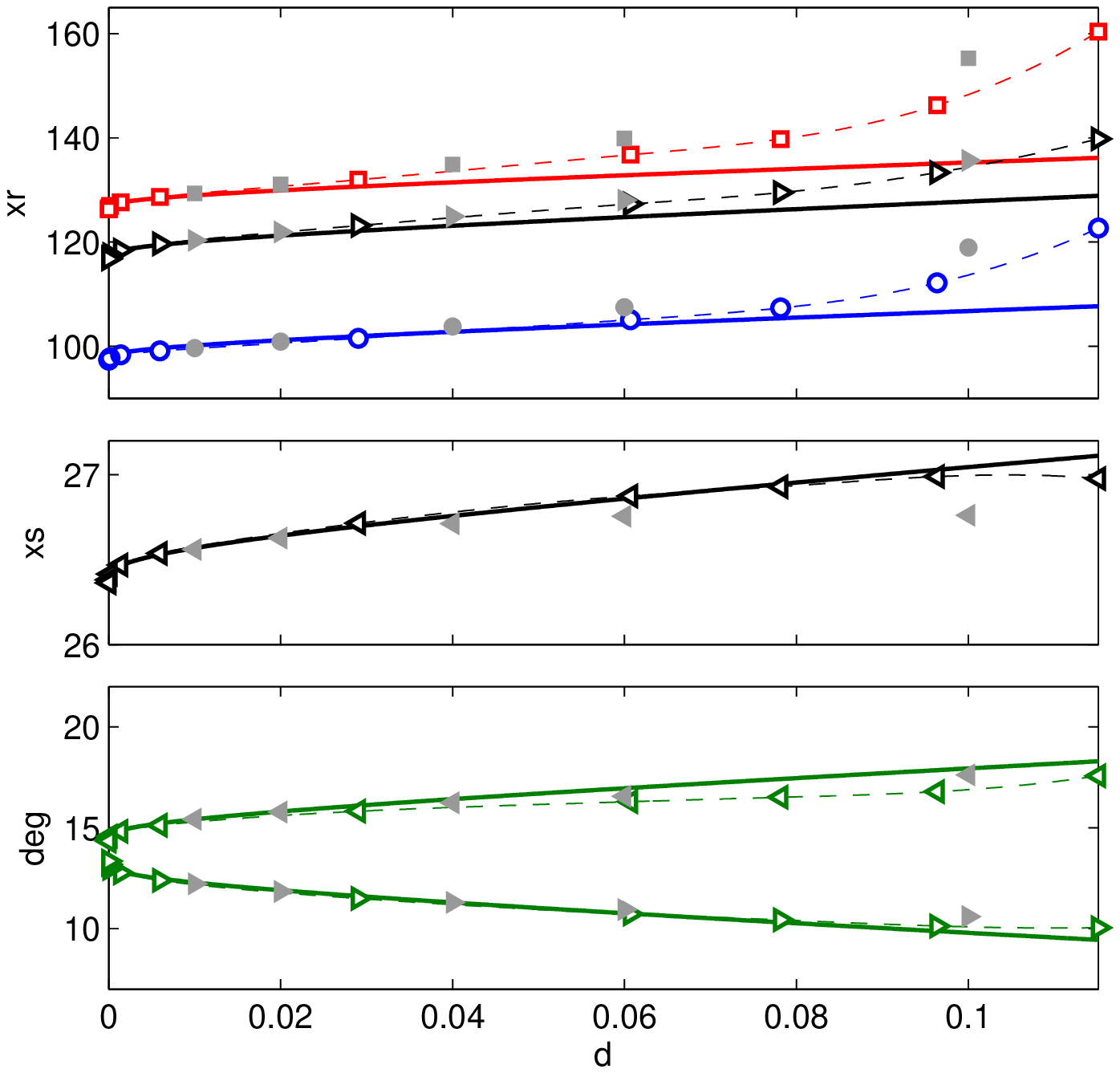}
	\put(-5, 91)  {$(a)$} 
   	\put(100, 91)  {\textcolor{red} {$A_{back}$ }} 
   	\put(100, 74)  {\textcolor{blue}{$A_{rec}/2$}} 	   	   	
   	\put(100, 82)  {$x_r$} 
   	\put( 15, 50)  {$x_s$} 
   	\put(18, 25)  {\textcolor[rgb]{0,0.5,0}{$\alpha_s$}} 	
   	\put(62, 9.5) {\textcolor[rgb]{0,0.5,0}{$180-\alpha_r$}}   
  	\end{overpic}
  }
  \vspace{0.5cm}
  \centerline{
	\psfrag{y}[r][][1][-90]{$y$}	
	\psfrag{x} [t][b]{$x$}
   	\begin{overpic}[width=9.5 cm,tics=10]{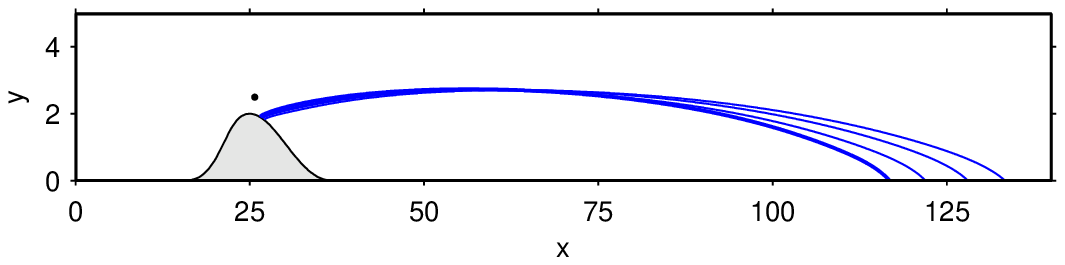}
   	 	\put(-5, 22) {$(b)$} 
  	\end{overpic}
  }
  \caption{Effect of a small control cylinder of diameter $d$ 
   inserted in the flow at $(x_c,y_c)=(25.7,2.5)$, at $\Rey=500$.
  $(a)$
  Variation of characteristic separation quantities (recirculation and backflow areas, stagnation locations, separation and reattachment angles).
  Thick solid lines show theoretical predictions from sensitivity analysis while open symbols show results from non-linear calculations, both with the force model (\ref{eq:SmallCylForce}).
  Grey filled symbols show results from non-linear calculations with the control cylinder included in the mesh. 
  $(b)$
  Separatrix for the uncontrolled flow (thick line) and for $d=0.02$, 0.04  and 0.06 (thin lines).}
   \label{fig:volume_control}
\end{figure}

Figure~\ref{fig:volume_control} shows variations of quantities of interest when a small control cylinder of diameter $d=0.05$ is inserted in the flow at $\xx_c=(x_c,y_c)=(25.7,2.5)$ (this location is shown in figure~\ref{fig:SmallCyl} with a black dot).
All quantities increase with the cylinder diameter, especially the reattachment point and recirculation and backflow areas, while the separatrix angles and  separation point are less affected.
Again, non-linear effects are observed when $d \gtrsim 0.05$.
This is consistent with the non-linear variations measured by Parezanovi\'{c} \& Cadot \cite{Par12} with $d=1$~mm and 3~mm.
Recall that in addition to the non-linear effects neglected by our linear sensitivity analysis, variations (\ref{eq:SmallCylEffectRL}) are linear in $\bdelta \CC$ but not in $d$ (since the drag coefficient depends on $d$), therefore sensitivity predictions in figure~\ref{fig:volume_control} are not straight lines, unlike those in figure~\ref{fig:wall_control}.
We also report  results obtained with the control cylinder  included in the computational mesh (grey filled symbols) in order to assess the assumption of uniform flow underlying (\ref{eq:SmallCylForce}).
Differences can be noticed when $d \gtrsim 0.05$, but sensitivity analysis does provide useful qualitative information regarding the influence of small passive control devices.

\section{Conclusion}

Considering the boundary layer flow above a wall-mounted bump as a prototype for separated flows, a variational technique was used to derive analytical expressions for the sensitivity of several geometric indicators of flow separation to steady actuation: 
the locations of the two stagnation points ($x_s$ and $x_r$), 
the  angles of the separatrix ($\alpha_s$ and $\alpha_r$) at these  points, 
the backflow      area $A_{back}$ and 
the recirculation area $A_{rec}$. 
For each geometric quantity, analytical expressions for the linear sensitivity to base flow modification $\nabla_{\UU} *$ were obtained. This gradient information was further translated in sensitivity maps to localised volume forcing and wall blowing/suction through the introduction of the adjoint base flow, governed by  linear adjoint equations forced by the previously determined gradient $\nabla_{\UU} *$. 
A suitable modelling of the addition of a small control cylinder as a localised force  depending on the local velocity allowed to obtain  sensitivity maps  relevant to experimental studies. 

Validations against full non-linear Navier$-$Stokes calculations showed an excellent agreement for small-amplitude control for all considered indicators. 
Non-linear effects appeared at larger amplitudes, consistent with experimental observations.
With very resemblant sensitivity maps, the reattachment point, the backflow and recirculation areas were seen to be easily manipulated. 
In contrast, the upstream separation point and the separation and reattachment angles were found to remain extremely robust with respect to external steady actuation. 

The present analysis, however, is limited to steady actuation and calls for a generalisation to 
the sensitivity of mean recirculation properties to harmonic forcing, which is known to be a more realistic, reliable and efficient experimental control scheme. 
Additionally, the recent development of fast imaging techniques has now made these geometric descriptors accessible in real-time \cite{Gau13a}, highlighting the need for the generalisation of  current open-loop   control optimisation tools to the dynamic closed-loop control of separation. 
The recent development of a solid theory for unsteady separation \cite{Hal04,Wel08} provides a firm ground for this challenging  objective.

\section*{Acknowledgment}

The authors thank Pierre-Yves Lagr\'ee for an interesting discussion about Lighthill's original derivation of the separatrix angle, and for pointing to an Eulerian definition of the separatrix.
This work was supported by the Swiss National Science Foundation (grant no. 200021-130315) and  the French National Research Agency (project no. ANR-09-SYSC-001).

\bibliographystyle{vancouver}
\bibliography{biblio_sep}

\begin{thebibliography}{10}

\bibitem{Seifert06}
Seifert A, Pack~Melton L.
\newblock Identification and control of turbulent boundary layer separation,
  G\"{o}ttingen, Germany, August 2004.
\newblock In: Meier GEA, Sreenivasan KR, Heinemann HJ, editors. IUTAM Symposium
  on One Hundred Years of Boundary Layer Research. New York, NY: Springer;
  2006. p. 199--208.

\bibitem{Alam06}
Alam MR, Liu W, Haller G.
\newblock Closed-loop separation control: An analytic approach.
\newblock Physics of Fluids (1994-present). 2006;18(4).
\newblock Available from: \url{http://dx.doi.org/10.1063/1.2188267}.

\bibitem{Hal04}
Haller G.
\newblock Exact theory of unsteady separation for two-dimensional flows.
\newblock Journal of Fluid Mechanics. 2004 8;512:257--311.
\newblock Available from:
  \url{http://journals.cambridge.org/article_S0022112004009929}.

\bibitem{Juillet13}
Juillet F, Schmid PJ, Huerre P.
\newblock Control of amplifier flows using subspace identification techniques.
\newblock Journal of Fluid Mechanics. 2013 6;725:522--565.
\newblock Available from:
  \url{http://journals.cambridge.org/article_S0022112013001948}.

\bibitem{hen07}
Henning L, King R.
\newblock Robust multivariable closed-loop control of a turbulent
  backward-facing step flow.
\newblock Journal of Aircraft. 2007;44(1):201--208.
\newblock Available from: \url{http://arc.aiaa.org/doi/abs/10.2514/1.22934}.

\bibitem{Gau13a}
Gautier N, Aider JL.
\newblock Control of the separated flow downstream of a backward-facing step
  using visual feedback.
\newblock Proceedings of the Royal Society A. 2013;469(2160).
\newblock Available from: \url{http://dx.doi.org/10.1098/rspa.2013.0404}.

\bibitem{Seifert96}
Seifert A, Darabi A, Wygnanski I.
\newblock Delay of airfoil stall by periodic excitation.
\newblock Journal of Aircraft. 1996;33:691--698.
\newblock Available from: \url{http://dx.doi.org/10.2514/3.47003}.

\bibitem{Garnier12}
Garnier E, Pamart PY, Dandois J, Sagaut P.
\newblock Evaluation of the unsteady {RANS} capabilities for separated flows
  control.
\newblock Computers \& Fluids. 2012;61:39--45.
\newblock Available from:
  \url{http://www.sciencedirect.com/science/article/pii/S0045793011002684}.

\bibitem{McLachlan89}
McLachlan BG.
\newblock Study of a circulation control airfoil with leading/trailing-edge
  blowing.
\newblock Journal of Aircraft. 1989;26(2):817--821.
\newblock Available from: \url{http://dx.doi.org/10.2514/3.45846}.

\bibitem{Fie90}
Fiedler HE, Fernholz HH.
\newblock On management and control of turbulent shear flows.
\newblock Progress in Aerospace Sciences. 1990;27(4):305 -- 387.
\newblock Available from:
  \url{http://www.sciencedirect.com/science/article/pii/0376042190900022}.

\bibitem{Wilson13}
Wilson J, Schatzman D, Arad E, Seifert A, Shtende T.
\newblock Suction and Pulsed-Blowing Flow Control Applied to an Axisymmetric
  Body.
\newblock AIAA Journal. 2013;51(10):2432--2446.
\newblock Available from: \url{http://dx.doi.org/doi:10.2514/1.J052333}.

\bibitem{Pujals10Exp}
Pujals G, Depardon S, Cossu C.
\newblock Drag reduction of a 3D bluff body using coherent streamwise streaks.
\newblock Experiments in Fluids. 2010;49(5):1085--1094.
\newblock Available from: \url{http://dx.doi.org/10.1007/s00348-010-0857-5}.

\bibitem{Par12}
Parezanovi\'c V, Cadot O.
\newblock Experimental sensitivity analysis of the global properties of a
  two-dimensional turbulent wake.
\newblock Journal of Fluid Mechanics. 2012 2;693:115--149.
\newblock Available from:
  \url{http://journals.cambridge.org/article_S0022112011004952}.

\bibitem{Wang03}
Wang Y, Haller G, Banaszuk A, Tadmor G.
\newblock Closed-loop Lagrangian separation control in a bluff body shear flow
  model.
\newblock Physics of Fluids (1994-present). 2003;15(8):2251--2266.
\newblock Available from:
  \url{http://scitation.aip.org/content/aip/journal/pof2/15/8/10.1063/1.1588636}.

\bibitem{Bou14}
Boujo E, Gallaire F.
\newblock Controlled reattachment in separated flows: a variational approach to
  recirculation length reduction.
\newblock Journal of Fluid Mechanics. 2014 3;742:618--635.
\newblock Available from:
  \url{http://journals.cambridge.org/article_S0022112014000238}.

\bibitem{Zie97}
Zielinska BJA, Goujon-Durand S, Du\v{s}ek J, Wesfreid JE.
\newblock Strongly Nonlinear Effect in Unstable Wakes.
\newblock Phys Rev Lett. 1997 Nov;79:3893--3896.
\newblock Available from:
  \url{http://link.aps.org/doi/10.1103/PhysRevLett.79.3893}.

\bibitem{choi99}
Choi H, Hinze M, Kunisch K.
\newblock Instantaneous control of backward-facing step flows.
\newblock Applied Numerical Mathematics. 1999;31(2):133 -- 158.
\newblock Available from:
  \url{http://www.sciencedirect.com/science/article/pii/S0168927498001317}.

\bibitem{Pas13}
Passaggia PY, Ehrenstein U.
\newblock Adjoint based optimization and control of a separated boundary-layer
  flow.
\newblock Euro J Mech B/Fluids. 2013;41:169--177.
\newblock Available from:
  \url{http://dx.doi.org/doi:10.1016/j.euromechflu.2013.01.006}.

\bibitem{Bou13}
Boujo E, Ehrenstein U, Gallaire F.
\newblock Open-loop control of noise amplification in a separated boundary
  layer flow.
\newblock Physics of Fluids. 2013;25(12).
\newblock Available from:
  \url{http://scitation.aip.org/content/aip/journal/pof2/25/12/10.1063/1.4846916}.

\bibitem{Tan56}
Taneda S.
\newblock Experimental investigation of the wakes behind cylinders and plates
  at low {R}eynolds numbers.
\newblock Journal of the Physical Society of Japan. 1956;11:302--307.
\newblock Available from: \url{http://dx.doi.org/doi:10.1143/JPSJ.11.302}.

\bibitem{Acr68}
Acrivos A, Leal LG, Snowden DD, Pan F.
\newblock Further experiments on steady separated flows past bluff objects.
\newblock Journal of Fluid Mechanics. 1968;34:25--48.
\newblock Available from: \url{http://dx.doi.org/10.1017/S0022112068001758}.

\bibitem{nis78}
Nishioka M, Sato H.
\newblock Mechanism of determination of the shedding frequency of vortices
  behind a cylinder at low Reynolds numbers.
\newblock Journal of Fluid Mechanics. 1978 11;89:49--60.
\newblock Available from:
  \url{http://journals.cambridge.org/article_S0022112078002451}.

\bibitem{bar02}
Barkley D, Gomes M~Gabriela M, Henderson RD.
\newblock Three-dimensional instability in flow over a backward-facing step.
\newblock Journal of Fluid Mechanics. 2002;473:167--190.
\newblock Available from: \url{http://dx.doi.org/10.1017/S002211200200232X}.

\bibitem{Mar03}
Marquillie M, Ehrenstein U.
\newblock On the onset of nonlinear oscillations in a separating boundary-layer
  flow.
\newblock Journal of Fluid Mechanics. 2003 8;490:169--188.
\newblock Available from:
  \url{http://journals.cambridge.org/article_S0022112003005287}.

\bibitem{Gia07}
Giannetti F, Luchini P.
\newblock Structural sensitivity of the first instability of the cylinder wake.
\newblock Journal of Fluid Mechanics. 2007;581:167--197.
\newblock Available from: \url{http://dx.doi.org/10.1017/S0022112007005654}.

\bibitem{Pas12}
Passaggia PY, Leweke T, Ehrenstein U.
\newblock Transverse instability and low-frequency flapping in incompressible
  separated boundary layer flows: an experimental study.
\newblock Journal of Fluid Mechanics. 2012;703:363--373.
\newblock Available from: \url{http://dx.doi.org/10.1017/jfm.2012.225}.

\bibitem{Ehr05}
Ehrenstein U, Gallaire F.
\newblock On two-dimensional temporal modes in spatially evolving open flows:
  The flat-plate boundary layer.
\newblock J Fluid Mech. 2005;536:209--218.
\newblock Available from:
  \url{http://dx.doi.org/doi:10.1017/S0022112005005112}.

\bibitem{Lig63}
Lighthill MJ.
\newblock Introduction. {B}oundary layer theory.
\newblock In: Rosenhead L, editor. Laminar Boundary Layers. Oxford University
  Press; 1963. p. 46--113.

\bibitem{Wel08}
Weldon M, Peacock T, Jacobs GB, Helu M, Haller G.
\newblock Experimental and numerical investigation of the kinematic theory of
  unsteady separation.
\newblock Journal of Fluid Mechanics. 2008 9;611:1--11.
\newblock Available from:
  \url{http://journals.cambridge.org/article_S0022112008002395}.

\end{thebibliography}

\end{document}